\documentclass[11pt]{article}
\usepackage{moriond,epsfig}

\def\Journal#1#2#3#4{{#1} {\bf #2}, #3 (#4)}

\def\NPA{{\em Nucl. Phys.} A}
\def\PLB{{\em Phys. Lett.} B}
\def\PRL{\em Phys. Rev. Lett.}
\def\PRC{{\em Phys. Rev.} C}
\def\PRD{{\em Phys. Rev.} D}

\def\EPC{{\em Eur. Phys. J.} C}
\def\ACS{\em Act. Phys. Slov}

\def\be{\begin{equation}}
\def\ee{\end{equation}}
\def\bea{\begin{eqnarray}}
\def\eea{\end{eqnarray}}

\begin{document}
\title{Hadron Correlations and Fluctuations in 40, 80, and 158 AGeV/$c$
Pb-Au Collisions}

\author{Hiroyuki Sako \\ for the CERES/NA45 Collaboration}

\address{Gesellschaft f\"ur Schwerionenforschung (GSI) \\ 64291 Darmstadt, Germany}

\maketitle\abstracts{
Measurements of HBT correlations and event-by-event fluctuations
of mean $p_T$ and the net charge
in Pb-Au collisions at 40, 80, and 158~AGeV/$c$ are presented.
From comparisons of HBT radii measured from AGS to RHIC energies,
a universal mean free path of pions at the thermal freeze-out
of about 1~fm is derived.
Non-statistical mean $p_T$ fluctuations of about 0.7~\% 
are measured, which are somewhat smaller than fluctuations at RHIC.
No indication for the QCD critical point is observed.
Fluctuations of the net charge are reproduced by {\sc RQMD} and {\sc
 URQMD} models, but significantly larger than prediction in
equilibrated QGP.
}

\section{Introduction}

The HBT interferometry measurements of pions provide information on
the space-time extent and the dynamical behavior
of the pion source in relativistic heavy ion collisions.
The present data by CERES at 40, 80, and 158 AGeV~\cite{Adamova:2002,Adamova:2003}
serve as an important link between the existing results from AGS, SPS,
and RHIC, which may provide a hint to the recently found
RHIC puzzle~\cite{larry}.
By systematic comparisons of the source size
among different collision energies, thermal freeze-out conditions may
be derived.

Event-by-event mean $p_{T}$ fluctuations
have been proposed as a probe to search
for the QCD critical point,
where the fluctuations are predicted to be enhanced~\cite{Stephanov:1998,Stephanov:1999}.
Fluctuations of the net electric charge
have been proposed as a probe to search for initial
fluctuations in the QGP phase, where the fluctuations
are predicted to be suppressed~\cite{Asakawa:2000,Jeon-Koch:2000}.
The present data may also provide a hint to answer the question whether initial suppressed fluctuations in
the QGP can survive hadronization and subsequent rescattering.

A detailed description of the CERES experiment is found
elsewhere~\cite{ceres-ptfluc-2003}. In the present analysis,
the primary vertex and charged particle tracks
are reconstructed by two silicon drift detectors (SDD's)
located just after the target, and a Time Projection Chamber (TPC)
located $\sim 4$~m downstream from the target.
The acceptance of a full-length track
in the TPC is $2.2 < \eta < 2.7$. Momentum of a charged
particle track is calculated from its azimuthal deflection
inside the TPC under the magnetic field.
The centrality of a Pb-Au collision is determined
by the charged particle multiplicity in SDD's at 40 AGeV/$c$, and
an analog signal in the multiplicity counter at 80 and
158 AGeV/$c$. The data analyzed here correspond to the $\sim 20$~\% most
central events.

\section{Results of HBT correlations}
\begin{figure}
\center
\epsfig{figure=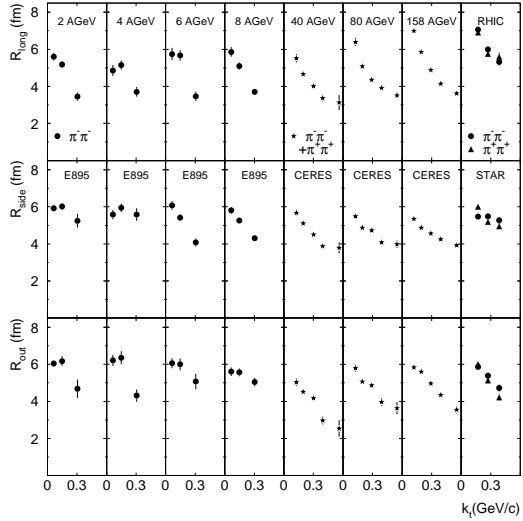,height=7cm}
\caption{Source radii (fm) as a function of the mean pair $p_{T}$ (GeV/c)
at different collision energies at AGS, SPS, and RHIC.}
\label{fig:hbt-s}
\end{figure}

%
Two identical pion correlation functions are fitted to a formula based
on the Bertsch-Pratt parameterization with the Cartesian decomposition of
$q = (q_{long},q_{side},q_{out})$
in the longitudinally comoving system (LCMS) assuming pion mass.
The fitting formula includes a consistent Coulomb
correction~\cite{Sinyukov}.
Fig.~\ref{fig:hbt-s} shows the three components of the source radii,
$R_{long}$, $R_{side}$, and $R_{out}$ as a function of mean pair
$p_{T}$ ($k_{T}$) at AGS, SPS, and RHIC.
The variations of the source radii in these collision energies are
small. The $k_{T}$ dependence shows a strong decrease.
At SPS energies, this dependence is shown to be described
with collective expansion models~\cite{Adamova:2002}.

$R_{long}$, proportional to the source life time, increases from AGS to RHIC energies.
$R_{side}$, corresponding to the transverse source size at small
pair transverse-mass, decreases from
AGS energy to 158 AGeV/$c$ at SPS, and then increases to the RHIC
energy. The ratio $R_{out}/R_{out}$ is observed to be close
to 1 at each collision energy, which may be related to
a short duration of particle emission, and possible opaqueness
of a pion source~\cite{vischer,borisflow}.

\begin{figure}
\epsfig{figure=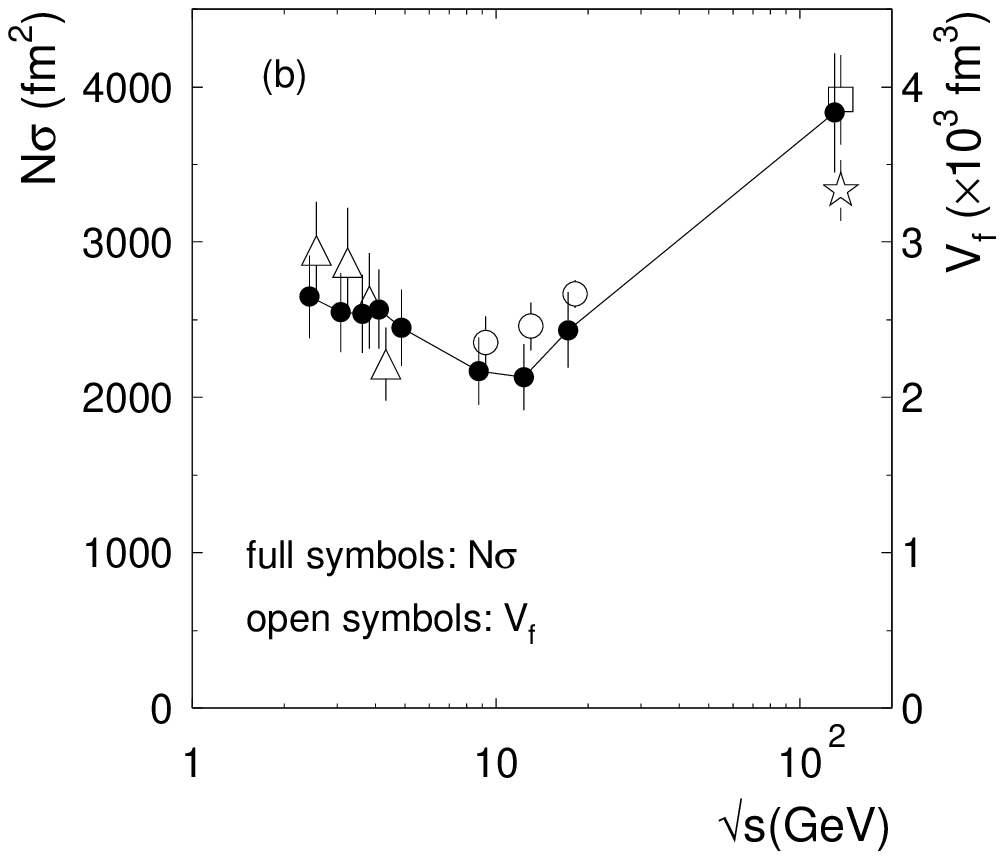,height=6.9cm}
\epsfig{figure=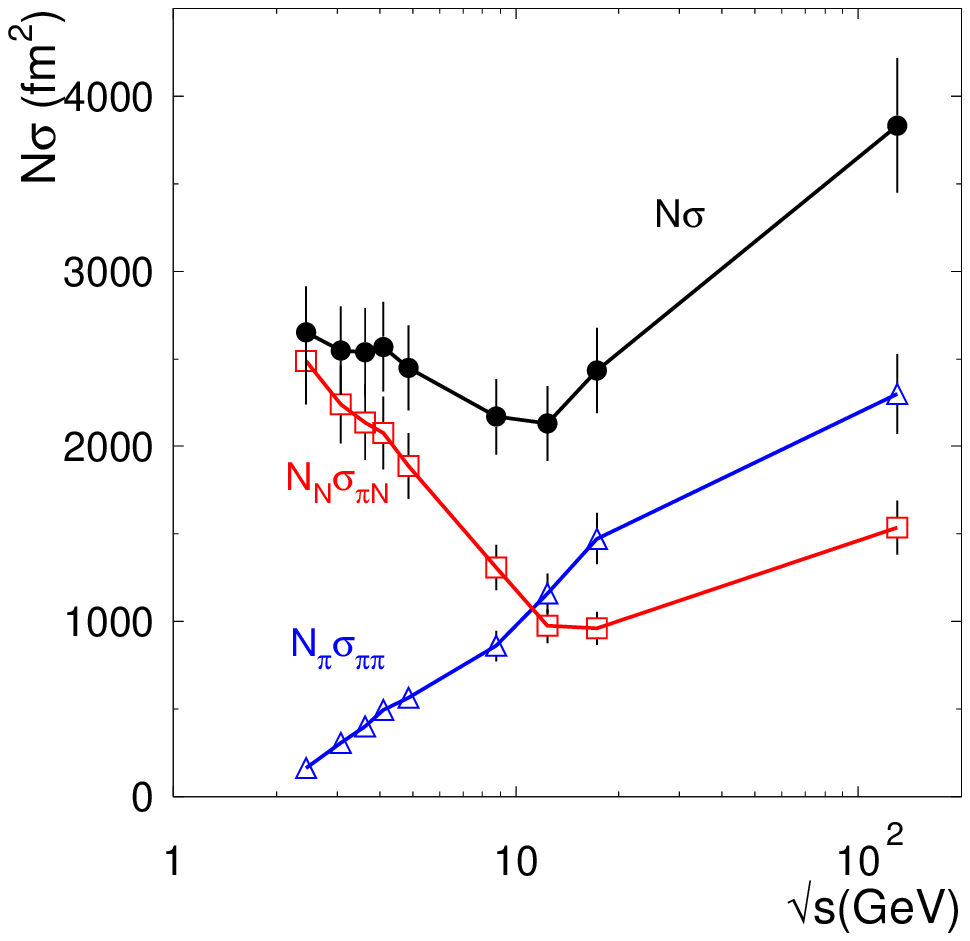,height=6.7cm}
\label{fig:hbt-mfp}
\caption{Left: the freeze-out volume $V_{f}$ (in the right y-axis) 
and $N \sigma$ (in the left y-axis) near mid-rapidity as a function of $\sqrt{s_{NN}}$.
Right: the $N \sigma$ of pion and nucleon components and the sum as a function
  of $\sqrt{s_{NN}}$.}
\end{figure}

The freeze-out volume is expressed as $V_{f} =
(2\pi)^{3/2}R_{long}R_{side}^2$ at small $k_{T}$.
The $V_{f}$ obtained from the measured source radii at $k_{T} \sim
0.16$ GeV$/c$ is plotted in the left panel
of Fig.~\ref{fig:hbt-mfp} as a function
of $\sqrt{s_{NN}}$~\cite{Adamova:2003}. The $V_{f}$ shows a non-monotonic
dependence, with a minimum between AGS and SPS energies.
To understand this, we consider $N \sigma$, where $N$ is
the number of the particles inside $V_{f}$, and $\sigma$ is the
cross section of a particle in the medium with a pion.
Neglecting particles except
for pions and nucleons, we can express nucleon and pion components
separately as $N \sigma = N_{N} \sigma_{\pi N} + N_{\pi} \sigma_{\pi\pi}$.
The two components of $N \sigma$ and the sum are plotted in the right
panel. The total $N\sigma$ shows again a non-monotonic dependence.
In the left panel~\cite{Adamova:2003}, the $N\sigma$ shows a very
similar collision energy dependence to $V_{f}$.
The scaling factor is given by the mean free path of pions
at the freeze-out, $\lambda_{f} = V_{f}/(N\sigma)$. Thus,
a universal freeze-out condition $\lambda_{f} \approx 1$~fm is derived
from AGS to RHIC energies.

\section{Results of event-by-event fluctuations}
\begin{figure}
\center
\epsfig{figure=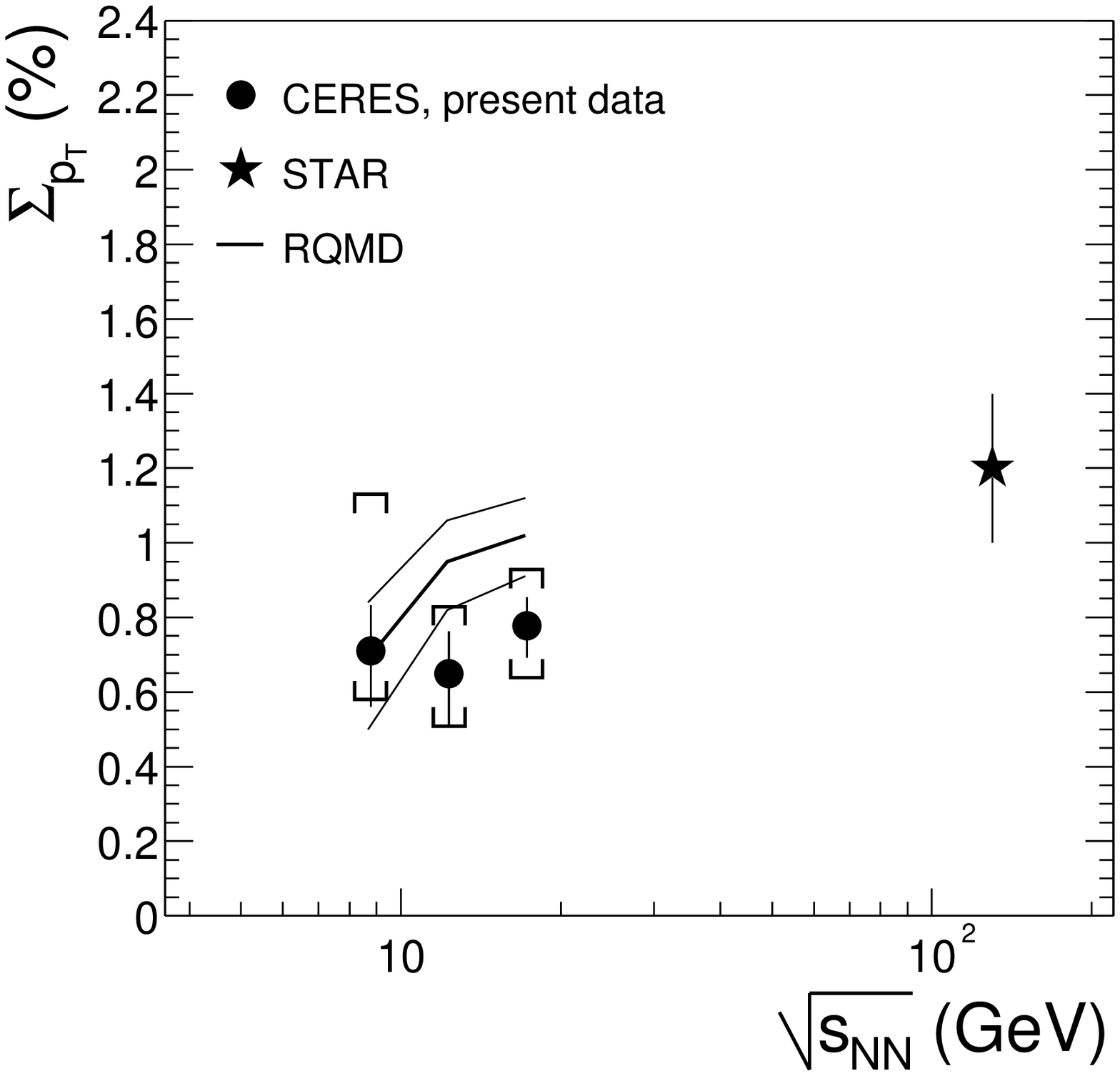,height=7.2cm}
\epsfig{figure=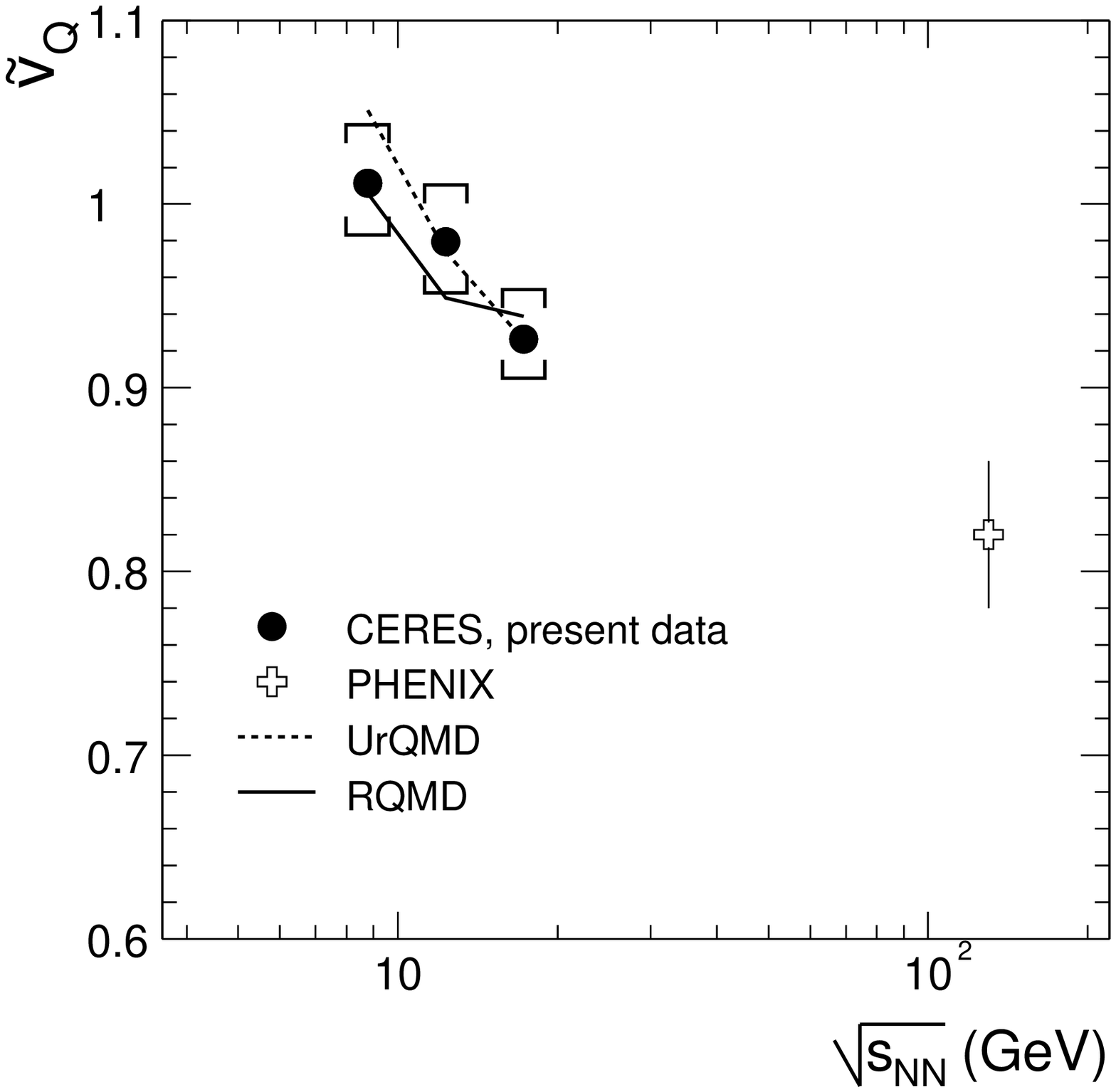,height=7cm}
\caption{Mean $p_{T}$ (left) and net charge (right) fluctuations as a
  function of $\sqrt{s_{NN}}$.}
\label{fig:ebe-fluct}
\end{figure}

A measure of mean $p_{T}$ fluctuations is defined as
$\Sigma_{p_{T}} \equiv
sgn(\sigma_{real}^2-\sigma_{stat}^2)\cdot
\frac{\sqrt{|\sigma_{real}^2-\sigma_{stat}^2|}}{\overline{p_{T}}}$,
where $\overline{p_{T}}$ is the mean $p_{T}$,
$\sigma_{real}$ is the r.m.s.~of the event-by-event mean $p_{T}$
distribution from the real events, and $\sigma_{stat}$ is that from
the statistical distribution, which is defined as $\sigma_{inc}/\sqrt{\langle
 N\rangle}$ with the r.m.s.~of the inclusive $p_{T}$ distribution,
$\sigma_{inc}$, and the mean multiplicity, $\langle N \rangle$~\cite{Voloshin:1999}.
The left panel of Fig.~\ref{fig:ebe-fluct} shows mean $p_{T}$
fluctuations measured by CERES at 40, 80, and 158~AGeV/$c$
in the 6.5~\% most central events
at $0.1<p_{T}<1.5$~GeV/$c$ and $2.2 < \eta < 2.7$,
which are compared to the measurement by STAR at $\sqrt{s_{NN}} = 130$~GeV in $0.1<p_{T}<2.0$~GeV/$c$~\cite{Voloshin:2001}.
The CERES data, corrected for short range $q$
correlations and two-track resolutions~\cite{ceres-ptfluc-2003}
are about 0.7~\%, which are
smaller than the uncorrected STAR data of about 1.2~\%.
No indication of non-monotonic
dependence from SPS to RHIC indicating the crossing of the QCD
critical point was observed.
The {\sc RQMD} model gives slightly higher fluctuations than the CERES data.

A measure of net charge fluctuations is defined as
$\tilde{v}_{Q} \equiv \frac{1}{C_{y}C_{\mu}}\frac{\langle \Delta Q^2
  \rangle}{\langle N_{+} \rangle + \langle N_{-} \rangle}$.
Here $\langle \Delta Q^2 \rangle$ is the r.m.s.~of the event-by-event net
charge ($Q \equiv N_{+} - N_{-}$) distribution from the real events,
where $N_{+}$ and $N_{-}$ are positive and negative charged particle
multiplicity, respectively.
Two correction factors are applied; $C_{y}$ for the
global charge conservation, and $C_{\mu}$ for the
non-zero mean net charge.
The right panel of Fig.~\ref{fig:ebe-fluct} shows the net charge
fluctuations measured by CERES at the three beam energies in the
6.5~\% most central events at $0.1<p_{T}<2.5$~GeV/$c$ and at
$2.05<\eta<2.85$, which are compared with the measurement by
PHENIX at RHIC~\cite{PHENIX-d:2002}.
A decrease of $\tilde{v}_{Q}$ from 1 to $\sim 0.85$ as a function of
$\sqrt{s}$ is observed from SPS to RHIC.
The observed fluctuations at SPS are reproduced well
by the {\sc RQMD} and {\sc URQMD} models,
while they are significantly underestimated by
QGP predictions of $\sim 0.3$. 

\section{Conclusions}
We measured pion source radii in 40, 80, and 158 AGeV/$c$ Pb-Au
collisions.
The collision energy dependence of
the freeze-out volume obtained with $R_{long}$ and $R_{side}$ shows
non-monotonic behavior with a minimum between AGS and SPS energies.
We derived a universal mean free path
of about 1~fm at the thermal freeze-out, which is independent of the collision energy. 
The small mean free path may indicate opaqueness
of the source, which may explain
also the observation of the smaller $R_{\rm out}$ than
$R_{\rm side}$ at SPS and RHIC~\cite{vischer,borisflow,padula}.

We measured mean $p_{T}$ fluctuations of about 0.7\%,
which are somewhat smaller than the measurement by STAR at 
$\sqrt{s_{NN}} = 130$~GeV.
The collision energy dependence does not show an
indication for the QCD critical point~\cite{Stephanov:1999}. 
The {\sc RQMD} model is slightly higher than the observed data at SPS energies.

The magnitude of net charge fluctuations
at 40, 80, and 158~AGeV/$c$ is significantly
larger than the expectation for an equilibrated QGP.
The good agreement of the observed fluctuations
with hadronic cascade models suggests
that resonance dynamics and hadronic diffusions are
the dominant sources of the observed fluctuations~\cite{Zaranek:2001}.

The diffusion size of hadrons in rapidity is
modeled as $\Delta y_{\rm diff} \sim \delta y_{\rm coll}
\sqrt{\frac{\tau_{\rm had}}{\tau_{\rm free}}}$~\cite{ShuSteph:2001}, where
$\delta y_{\rm coll}$ is a mean rapidity shift per collision,
$\tau_{\rm had}$ is the life
time of the hadronic phase, and $\tau_{\rm free}$ is the mean free
time. The two time scales are expected to be much smaller than the
source size from the HBT measurements, and they affect the
diffusion size oppositely. It is therefore a key to evaluate those time scales
precisely to search for the onset of the QGP.

\section*{Acknowledgments}
This work was supported by the German BMBF,
the U.S.~DoE, the Israeli Science Foundation, and the MINERVA Foundation.


\end{document}